\documentclass[%
 reprint,
 amsmath,amssymb,
 aps,
]{revtex4-1}
\usepackage{graphicx}
\usepackage{caption}
\usepackage{subcaption}
\usepackage{amsthm,amsmath}
\usepackage[top=25.4mm, bottom=20.9mm, left=19.1mm, right=19.1mm]{geometry}
\usepackage[font=footnotesize,labelfont=bf]{caption}
\theoremstyle{definition} % age in karo nakonim az jayi ke theorem(inja example) estefade konim, ta akare file fonte hame kharab mishe

\newcommand{\DD}{\, \displaystyle}

\newcommand{\fracc}[2]{\, \displaystyle \frac{ #1}{ #2}}

\newcommand{\BB}[2]{\, {#1 \choose #2}}
\newcommand{\ST}[2]{\, {#1 \brack #2}}
\newcommand{\morabba}[1]{\,\begin{flushright}
 \Rectsteel \\
\end{flushright}}

\newcommand{\all}[2]{\,\begin{align}
                   #1 
                    \label{#2}
                   \end{align}
}

\makeatletter
\newcommand{\vast}{\bBigg@{4}}
\newcommand{\Vast}{\bBigg@{5}}

\makeatother

\begin{document}
\preprint{APS/123-QED}
%\date{}
%opening
\title{Multiplex Networks with Intrinsic Fitness: \\ 
 Modeling the Merit-Fame Interplay via Latent Layers}

\author{Babak Fotouhi$^{1}$ and Naghmeh Momeni$^{2}$ \\
$^{1}$Department of Sociology\\
$^{2}$Department of Electrical and Computer Engineering\\
McGill University, Montr\'eal, Qu\'ebec, Canada\\
Email:\texttt{ babak.fotouhi@mail.mcgill.ca}}

%\newgeometry{top=19.1mm, bottom=19.1mm, left=19.1mm, right=19.1mm}

\begin{abstract}
 \textbf{We consider the problem of growing multiplex networks with intrinsic fitness and inter-layer coupling. The model comprises two layers;  one that  incorporates fitness and another in which attachments are preferential. In the first layer, attachment probabilities are proportional to fitness values, and in the second layer, proportional to the sum of degrees in both layers.   
We provide analytical closed-form  solutions for the joint distributions of fitness and degrees.  We also derive closed-form expressions for the expected value of the degree as a function of fitness. The model alleviates two shortcomings that are present in the current models of growing multiplex networks: homogeneity of connections, and homogeneity of fitness. In this paper, we posit and analyze a  growth model that is heterogeneous  in both senses. 
}
\end{abstract}

%\author{Babak Fotouhi, Michael Rabbat}
 \maketitle
 \emph{keywords:} multiplex networks, intrinsic fitness, growing networks, preferential attachment, Stirling numbers. 

%\newgeometry{top=19.1mm, bottom=19.1mm, left=19.1mm, right=19.1mm}

\section{Introduction}

Multiplex networks are mathematical  tools for modeling   systems with multiple types of interaction. The system is conceptualized as being comprised of multiple layers, each hosting a distinct type of link (which corresponds to a type of interaction) between nodes. The set of nodes are the same for all layers. 
 Many real systems have been  modeled  under the  multiplex framework, such as
  citation networks~\cite{citation_aps,aps_imdb}, online social media~\cite{online}, airline networks~\cite{air}, scientific  collaboration  networks~\cite{citation_aps}, urban transportation networks~\cite{arenas0}, and online games~\cite{game}. Theoretically, multiplex networks demonstrate how incorporating additional dimensions and types of interaction to simple one-layer systems can change their dynamics, add new properties and alter existing ones. Diverse processes have been studied  theoretically on multiplex networks.   Examples include epidemics~\cite{epid1,epid2}, pathogen-awareness interplay~\cite{interplay}, percolation processes~\cite{percolation1,percolation2,epid1}, evolution of cooperation~\cite{cooperation1,cooperation2}, diffusion processes~\cite{diffusion} and social contagion~\cite{information}. For  thorough reviews, see~\cite{survey,matjaz}.

In the present paper we focus on the problem of growing multiplex networks with fitness.   
Previous studies on growing multiplex networks   exhibit  two main shortcomings:  homogeneity of growth, and homogeneity (or absence) of nodal  fitness. 
In~\cite{Bian1}, a growing two-layer network is studied, and various attachment kernels are envisaged.   The number of links established by each newcomer is considered to be the same for both layers (In the Supplemental Material of~\cite{Bian1}, the possibility of heterogeneous growth rates is entertained in the asymptotic mean-field analysis of degrees of individual nodes within each layer.  However, the  effect of growth heterogeneity on  the single-layer and inter-layer degree distributions remains unknown).  Similarly, the model posited  and thoroughly analyzed  in~\cite{Bian2} exhibits homogeneity in the sense that, for a given node, the expected degree is the same across layers.   In real systems,   the nature of the connections in different  layers differ, since they pertain to distinct types of interaction.    For example, in~\cite{game}, the interactions between the players of a massive online game is mapped onto six distinct layers, and their average degrees are different. It would be plausible to devise a growth model which incorporates  heterogeneity  explicitly.   In~\cite{nameni}, the problem of homogeneity  is alleviated by considering heterogeneous link growth rates. In the present paper, we consider heterogeneous link growth rates. 
%In addition to  heterogeneity of connections, the presented   growth mechanism   incorporates inter-layer dependence. 
% To our knowledge, no model exists in the literature that incorporates dependence,  and closed-form expressions for degree distributions are obtained. 

Another unrealistic assumption that is made by the previous studies on growing multiplex networks is that the probability  for  each existing node  to receive links from incoming nodes  only depends on their degrees. For example, if we consider the network of citations between scientific papers, this assumption would mean that the inherent quality and novelty of the papers have no role in the future number of citations that they would receive. That is, only fame drives scientific success, not quality: when scholars cite a paper,  they  only    take into account  the number of citations that an existing paper has. This is obviously not the case. Similarly, consider the case of online social networks such as Twitter, Instagram, Pinterest, Google$+$ and Tumblr. In all these networks, each user can `follow'  other users. 
Assuming that links are established only based on existing degrees---and not incorporating any intrinsic fitness for the nodes---would be synonymous with disregarding the role of the quality of the content produced by each user on  her/his popularity. True that after a user becomes famous, the fame on its own contributes to further accumulation of followers (which is the rationale behind all preferential attachment models), but quality also has an undeniable role---especially, at the initial stages of the lifetime of each node (user). This motivates us to consider intrinsic fitness for nodes. In~\cite{fit_0,fit_1,fit_2,fit_3,fit_4}, intrinsic fitness is envisaged in the case of single-layer networks. To our knowledge, no fitness-based model on multiplex networks exists in the literature. 

We consider a growing directed multiplex network that comprises two layers. Each node is assigned an intrinsic fitness, which models  its quality.  The fitness of a node never changes.  Each node belongs to two layers: a \emph{merit} layer and a \emph{ fame} layer. In the former, fitness values are the sole drivers of the growth mechanism. In the fame layer,   attachment is preferential, that is, the probability that a node receives a link from a newcomer is proportional to the \emph{total} degree of that node, i.e., the sum of its degrees in both layers. For example, in the case of citation networks, the interpretation of the model is as follows. Two distinct types of citations can be discerned. The first type---the \emph{meritocratic} type---is when a scholar reads a paper, and cites it because of its content (a citation which would be given regardless of the number of citations that paper already has). Another type of citation is what we call \emph{fame-driven}. A paper can become trendy, or well-known in some literature (particularly true for seminal papers which initiate a new subfield), and many citations that it receives would be solely due to its fame---i.e., current number of citations, which itself is the total of meritocratic  and fame-based citations. For example, after a seminal paper initiates or revives a scientific domain, after the domain passes its inchoate stages, many papers will be remote from those seminal papers, but will still cite it because those papers are famous, not because their content is being used (even tangentially) in the new paper being published. It is imperative to note that  to  a scholar who wants to cite an existing paper, quality is \emph{latent}. That is, only the total number of links is observed;  fame-based and meritocratic citations are not distinguishable for the new incoming node (the new paper).  What is observable is   the collapsed network, in which the links are aggregated into one layer. 

Our model emulates the said merit-fame interplay. We focus on the interlayer joint distribution of degrees and fitness. We find $P(k,\ell,\theta)$, which is the (asymptotic)  fraction of nodes with fitness $\theta$ who have degree $k$ in the merit layer and degree $\ell$ in the fame layer. This is presented in Equation~\eqref{P_fin}. We also find $P(q,\theta)$, which is the fraction of nodes with fitness $\theta$ whose total degree is $q$. This is given in Equation~\eqref{Pq_fin}. The results depend on the distribution of fitness values, as well as the initial number of links that each new node emanates in each of the  layers. We also find the conditional expected total degree of nodes. That is, for a given fitness value, we find the expected number of total links. This  result is  presented  in Equation~\eqref{qbar}. 

The rest of the paper is organized as follows. After introducing notation and terminology, we describe the growth mechanism quantitatively. We then undertake the rate equation approach to quantify the evolution of $P(k,\ell,\theta)$ as a function of time. We then focus on the steady-state, when transients vanish, and  solve the resulting equations. We then  obtain  $P(q,\theta)$ through a straightforward transformation, and then use  it to find the conditional expected value of total degree.

%(that is, the number of links that each newly-born node establishes is the same for both  layers)
%according to preferential attachment is considered, and it is shown that $\overline{\ell}(k)$ (which is the average layer-2 degree of nodes whose layer-1 degree is $k$) is a function of $k$. 

%In~\cite{Bian1}, the number of links that each new incoming node establishes in both layers is equal to $m$. 

%Previous results on growing multiplex networks are confined to homogeneously-growing layers~\cite{survey,Bian2,Bian1}. In the present paper, we consider heterogeneously-growing layers: each incoming node establishes $\beta_1$ links in layer 1 and $\beta_2$ links in layer 2. We also solve the problem for the case where growth is uniform, rather than  preferential. We demonstrate that, surprisingly, the expression for $\overline{\ell}(k)$ is identical to that of the preferential case. We verify the theoretical  findings  with Monte Carlo simulations. 

\section{Notation and Terminology}

%Throughout the present paper, we will use the following convention about binomial coefficients. We assume that $\CC{a}{b}$ is zero for negative $a$ or $b$. Also, for summations we use the following notation:  an unbounded sum means that the summation is from $-\infty$ up to $+\infty$ (that is, ${\sum_j a_j = \sum_{j=-\infty}^{\infty} a_j}$). If the sum is bounded, the summation bounds will be explicitly written. 

The network is directed, and we use the terms degree and in-degree interchangeably. 
 Form node $x$, the fitness value is denoted by $\theta_x$.  The probability  distribution of fitness values is denoted by $\rho(\theta)$. The  layer-1 degree of node  $x$ is denoted by $k_x$, and the layer-2  degree of node $x$ is denoted by $\ell_x$. The total number of links of node $x$ is denoted by $q_x$, that is, ${q_x=\ell_x+k_x}$. 
 If a quantity depends on time, we will explicitly mention it. If time dependence is not mentioned, the steady-state value of the quantity is meant. For example, $k_x(t)$ is the degree of node $x$ at time $t$, and $k_x$ is the degree of node $x$ in the steady state, that is, in the limit as ${t \rightarrow \infty}$.

\section{Model}

The system initially comprises  $N(0)$ nodes, each with  two types of links. The links are assumed to be established on two separate layers, layer 1 (the merit layer)  and layer 2 (the fame layer).  Let us emphasize that layers embody the  set of nodes, but the sets of links differ. Suppose that there are $L_1(0)$ links in the first layer and $L_2(0)$ links in the second layer at the outset. 

%A quantity that pertains to layer $i$ will be addressed  by the prefix  \emph{layer-$i$}. For example, by  the term   ` layer-1 neighbors of node $x$' we mean the set of nodes that are adjacent to node $x$ in layer 1. Similarly, a layer-2 link is a link that resides in the second layer. 

The network grows by the successive addition of new nodes. Time increments in discreet steps, and at each timestep one new node is added to the network.  Each incoming node establishes $\beta_1$ layer-1 links and $\beta_2$ layer-2 links to the existing nodes.

Upon being born, the fitness of an incoming node is drawn from $\rho(\theta)$ and stays the same thereafter. 
The mean value of the fitness distribution, that is, the expected value of the fitness of incoming nodes, is denoted by $\mu$. 
% (see~\footnote{This assumption does not diminish the generality of the final results--- they hold for general non-integer values of $\theta$ as wel. The only modification to the calculations will be that every generating function (Z-transform) over $\theta$, that will emerge in the calculations below, will be replaced by the continuous analog of the Z-transform, that is, the Laplace transform. }). 

%In all the models discussed below, layer 1 has the additional trait compared to layer 2 that, the probabilities of receving a link in layer 1 depends on the fitness of nodes. By the term `a node of ${(k,\theta,\ell)}$'  we mean a node that has layer-1 degree $k$, fitness value $\theta$ and layer-2 degree $\ell$. 

%Throughout the paper, we will denote the number of nodes that have fitness $\theta$, layer-1 degree $k$ and layer-2 degree $\ell$ at time $t$  by ${N_t(k,\theta,\ell)}$. Also ${P_t(k,\theta,\ell)}$ fraction of nodes that has these properties, and  ${P(k,\ell,\theta)}$ represents the steady-state value of this fraction. 

\begin{figure}[h]
  \centering
  \includegraphics[width=.9 \columnwidth]{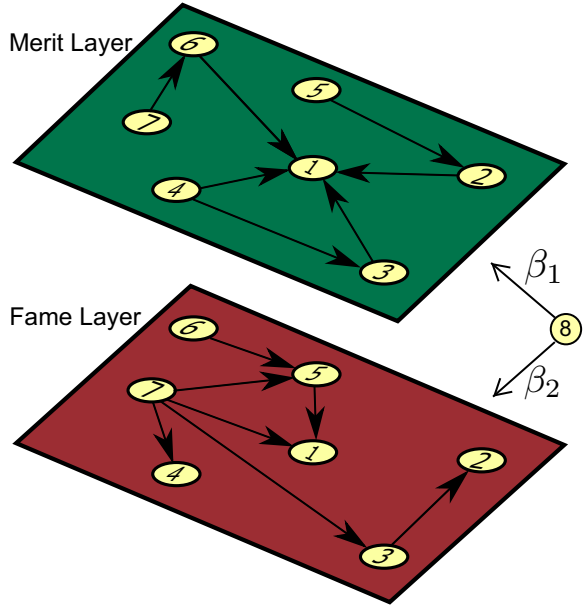}
  \caption[Figure ]%
  {\textbf Schematic illustration of the growth mechanism. A set of 7 nodes exist in the network, and node 8 is being appended. In the  merit layer, only the fitness values drive nodes' chances of receiving links from node 8 (who establishes $\beta_1$ links in this layer). In  the fame layer, the probability of receiving a link is proportional to total degree. For example, node 1 will have the highest chance of receiving a  link, because it has the greatest total degree (which is equal to six). Conversely, node 7  has the smallest total degree (which is zero), and will not get a link from node 8 in the fame layer. } 
\label{net}
\end{figure}

In the first layer, the probability of receiving a link from an incoming node for node $x$  is proportional to ${ \theta_x}$, where $\theta_x$ is the fitness of node $x$. In the second layer, the probability of node $x$ receiving a link from the newcomer is proportional to $k_x+\ell_x$. The probability of receiving a link in layer 1 can be written as ${\frac{ \theta_x}{\sum_x ( \theta_x)}}$. The sum in the denominator can be computed at time $t$ as follows. If the sum of the fitness values of the nodes at time $t=0$ is $\Theta$, then as time progresses, the sum of the fitness values of nodes converges to $\Theta+\mu t$, where $\mu$ is the mean of the fitness distribution. Since we will eventually limit the analysis to the steady state, the error of this approximation vanishes. For layer 2, the probability of receiving a link for node $x$ is equal to 
 ${\frac{ k_x+\ell_x}{\sum_x ( k_x+\ell_x)}}$. The sum in the denominator at any time equals the total number of links in both layers.%can be computed at time $t$ as follows.% The sum over $k_x$ yields    the total number of links. Denoting the number of links in the initial network by $L_0$, the denominator becomes $L_0+  \beta_1 t$. 
 %For the sum over fitness values, we will replace approximate individual fitness values by the average fitness $\mu$. The accuracy of this approximation improves with time. So the probability of link reception in layer 1 equals $\frac{k_x+\theta_x}{L_1(0)+(2\beta_1+\mu)t}$. 

\section{Joint Interlayer Distribution  of Degrees and Fitness }

At time $t$, upon the addition of the new node, certain events can change the value of ${N_t(k,\ell,\theta)}$. If a node  of ${(k-1,\ell,\theta)}$ receives a layer-1 link, then  it becomes a ${(k,\ell,\theta)}$ node, and  ${N_t(k,\ell,\theta)}$ increments consequently. 
 Similarly,   if a node  of ${(k,\ell-1,\theta)}$ receives a layer-2 link, then  it becomes a ${(k,\ell,\theta)}$ node, and  ${N_t(k,\ell,\theta)}$ increments consequently. On the other hand, if a node is already a ${(k,\ell,\theta)}$ node and it receives a link in either layer, it will no longer be a ${(k,\ell,\theta)}$ node, and ${N_t(k,\ell,\theta)}$ decrements consequently. Also, note that  the  layer-1 degree  and layer-2 degree  of each incoming node is 0 upon introduction, and such a node  has  fitness $\theta$ with probability $\rho(\theta)$. The following rate equation summarizes these events with their respective probabilities of occurrence, where $E \{\cdot\}$ denotes expected value:
 \all{
&E \{ N_{t+1}(k,\theta,\ell) \} = N_t(k,\theta,\ell)  
 \nonumber \\
 &
+
\beta_1  \fracc{\theta N_t(k-1,\ell,\theta)- \theta  N_t(k,\theta,\ell) }{\Theta+\mu t}
\nonumber \\
&
+
 \beta_2  \fracc{(k+\ell-1) N_t(k,\ell-1,\theta)- (k+\ell)  N_t(k,\theta,\ell) }{L_1(0)+L_2(0)+(  \beta_1 +\beta_2) t}
\nonumber \\ 
& +
\delta_{k 0} \delta_{\ell 0} \rho(\theta)
 .}{rate_1}

Hereinafter, we drop the expected value operator, and all the ${P(k,\ell,\theta)}$ numbers denote  expected values. Using the relation ${N_t(k,\theta,\ell)=(N(0)+t) P_t(k,\theta,\ell)}$,  we can rewrite~\eqref{rate_1} to quantify the evolution of ${P_t(k,\theta,\ell)}$ as follows:
 \all{
& \big[ N(0)+t \big] \big[ P_{t+1}(k,\theta,\ell) - P_t(k,\theta,\ell) \big] 
+ P_{t+1}(k,\theta,\ell) 
 = 
 \nonumber \\
 &
\beta_1  \fracc{\theta N_t(k-1,\ell,\theta)- \theta  N_t(k,\theta,\ell) }{\Theta+\mu t}
\nonumber \\
&
+
 \beta_2  \fracc{(k+\ell-1) N_t(k,\ell-1,\theta)-(k+\ell)  N_t(k,\theta,\ell) }{L_1(0)+L_2(0)+(  \beta_1 +\beta_2) t}
\nonumber \\ 
&+
\delta_{k 0} \delta_{\ell 0} \rho(\theta)
 .}{rate_2}

Note that negative $k$ or $\ell$ does not have a physical meaning, so  $k$ and $\ell$ in equation~\eqref{rate_2}, as well as every equation henceforth, are  nonnegative integers. 

Now we focus on the steady state, where by definition,  the values of  ${P_t(k,\theta,\ell)}$ reach horizontal asymptotes and their variations vanish. Also we note that in the limit as ${t \rightarrow \infty}$, we have
\all{
\begin{cases}
\DD \lim_{t \rightarrow \infty} \beta_1 \frac{N(0)+t}{\Theta+\mu  t} = \fracc{\beta_1}{\mu} \\  
\DD \lim_{t \rightarrow \infty} \beta_2  \frac{N(0)+t}{L_1(0)+L_2(0)+(  \beta_1 +\beta_2) t} = \fracc{\beta_2}{\beta_1+\beta_2} .
\end{cases}
}{limits_1}

Using these limits, we can rewrite~\eqref{rate_2} for the steady state as follows
\all{
&P(k,\ell,\theta) = \fracc{\beta_1}{\mu} \theta \bigg[   P(k-1,\ell,\theta) -  P(k,\ell,\theta)\bigg]
\nonumber \\
&
+ \frac{ \beta_2}{\beta_1+\beta_2} \bigg[(k+\ell-1)  P(k,\ell-1,\theta) - (k+\ell) P(k,\ell,\theta)  \bigg] 
\nonumber \\ &+ 
\delta_{k 0} \delta_{\ell 0} \rho(\theta)
.}{difference_1}

%\all{
%&P(k,\ell,\theta) \Bigg[1+\frac{\beta_1\theta}{\mu}+\frac{\beta_2}{\beta_1+\beta_2} (k+\ell) \Bigg]
%= \fracc{\beta_1}{\mu} \theta     P(k-1,\ell,\theta) 
%\nonumber \\
%&
%+ \frac{ \beta_2}{\beta_1+\beta_2}  (k+\ell-1)  P(k,\ell-1,\theta) 
%+
%\delta_{k 0} \delta_{\ell 0} \rho(\theta)
%.}{difference_2}
This can be rearranged and recast as

\all{
&P(k,\ell,\theta) \Bigg[\left(1+\frac{\beta_1\theta}{\mu}\right)\left(\frac{\beta_1+\beta_2}{\beta_2}\right)+k+\ell  \Bigg]
= \nonumber \\
&
\fracc{\beta_1(\beta_1+\beta_2)\theta}{\beta_2\mu}      P(k-1,\ell,\theta) 
+  (k+\ell-1)  P(k,\ell-1,\theta)  \nonumber \\ 
&
+
 \frac{\beta_1+\beta_2}{\beta_2}\delta_{k 0} \delta_{\ell 0} \rho(\theta) 
.}{difference_2}

Dividing both sides by the factor on the left hand side, this transforms into

\all{
&P(k,\ell,\theta)  
= \nonumber \\
&
\left(\fracc{\beta_1(\beta_1+\beta_2)\theta}{\beta_2\mu} \right)  \fracc{   P(k-1,\ell,\theta) }{\left(1+\frac{\beta_1\theta}{\mu}\right)\left(\frac{\beta_1+\beta_2}{\beta_2}\right)+k+\ell } \nonumber \\ &
+ \fracc{ (k+\ell-1)  P(k,\ell-1,\theta) }{\left(1+\frac{\beta_1\theta}{\mu}\right)\left(\frac{\beta_1+\beta_2}{\beta_2}\right)+k+\ell } \nonumber \\ 
&
+
 \fracc{\frac{\beta_1+\beta_2}{\beta_2}\delta_{k 0} \delta_{\ell 0} \rho(\theta) }{\left(1+\frac{\beta_1\theta}{\mu}\right)\left(\frac{\beta_1+\beta_2}{\beta_2}\right) +0+0}
.}{difference_2}

Hereinafter,  for brevity of notation, we denote  $\frac{(\beta_1+\beta_2) }{\beta_2 } $ by $A$,
% $\left(1+\frac{\beta_1\theta}{\mu}\right)\left(\frac{\beta_1+\beta_2}{\beta_2}\right) $ by $A$,
and we denote $\left(1+\frac{\beta_1\theta}{\mu}\right) $ by $G_{\theta}$. Thus the difference equation we need to solve takes the following form
 
\all{
P(k,\ell,\theta)  
= & 
\frac{A \theta  \beta_1}{\mu}  \fracc{   P(k-1,\ell,\theta) }{AG_{\theta}+k+\ell } \nonumber \\ &
+ \fracc{ (k+\ell-1)  P(k,\ell-1,\theta) }{AG_{\theta}+k+\ell }
+
 \fracc{ \delta_{k 0} \delta_{\ell 0} \rho(\theta) }{G_{\theta} }
.}{difference_4}

. 
 Let us define
 \all{
 \psi(k,\ell,\theta) \stackrel{\text{def}}{=} 
\fracc{\Gamma (AG_{\theta}  +k+\ell +1)}{\left(\frac{A\theta \beta_1}{\mu}\right)^k}
P(k,\ell,\theta)
 .}{psi_def}
 
 It is easy to verify the following relations using the properties of the Gamma function: 
 \all{
 \begin{cases}
\fracc{A \theta  \beta_1}{\mu}  \fracc{   P(k-1,\ell,\theta) }{A G_{\theta} +k+\ell }
=\fracc{\psi(k-1,\ell,\theta)}{\left(\frac{\mu}{A\theta \beta_1}\right)^k\Gamma  (A G_{\theta}  +k+\ell +1) }
 \\ \\
 \fracc{   P(k,\ell-1,\theta) }{A G_{\theta} +k+\ell }
=\fracc{\psi(k,\ell-1,\theta)}{ \left(\frac{\mu}{A\theta \beta_1}\right)^k\Gamma (A G_{\theta}  +k+\ell +1 )}
 .\end{cases}
 }{ms}

We substitute the first two terms on the right hand side of~\eqref{difference_4} with the expressions given in~\eqref{ms}. Then we multiply both sides by the factor ${\left(\frac{\mu}{A\theta \beta_1}\right)^k\Gamma (A G_{\theta}  +k+\ell +1 )}$. 
We arrive at the following difference equation: 
\all{
\psi(k,\ell,\theta)=&  \psi(k-1,\ell,\theta)+(k+\ell-1)\psi(k,\ell-1,\theta)
\nonumber \\ & + A\delta_{k 0} \delta_{\ell 0} \rho(\theta)
\Gamma (A G_{\theta})
.}{psi_diff}

Without loss of generality, we can take $k,k+\ell$ to be the arguments of the function instead of $k,\ell$. Let us define the new auxiliary function: 
\all{
\phi^{k+\ell}_{k}(\theta) \stackrel{\text{def}}{=} \psi(k,\ell,\theta)
.}{phi_def}
 (Note that $k+\ell$ is an upper index, not a power.) We can readily rewrite~\eqref{psi_diff} in terms of $\phi$. The difference equation reads
\all{
\phi^{k+\ell}_{k}(\theta)=&  \phi^{k+\ell-1}_{k-1}(\theta)
 +(k+\ell-1)\phi^{k+\ell-1}_k(\theta)
\nonumber \\ & + A\delta_{k 0} \delta_{\ell 0} \rho(\theta)
\Gamma (A G_{\theta})
.}{phi_diff}

Note that the last term on the right hand side is merely the boundary condition at $\{k,\ell\}=\{0,0\}$, and vanishes for any other combination of $k,\ell$. As our last change of variables, let us denote $k+\ell-1$ by $n$.  Dropping the $\theta$ argument for notational brevity, we can rewrite~\eqref{phi_diff} as a function of $n,k$ (without the term which dictates the  boundary condition at $\{k,\ell\}=\{0,0\}$) in the following form: 
\all{
\phi^{n+1}_k =\phi^{n}_{k-1}+n\phi^{n}_k.
}{stirling}
 
 This is the recurrence relation which defines the unsigned  Stirling numbers of the first kind. We denote the Stirling numbers by ${n\brack m}$ in this paper. Incorporating the initial conditions, the solution to~\eqref{stirling} is
 \all{
 \phi^{n}_k=\ST{n}{k} \times A \Gamma(AG_{\theta}) \rho(\theta)
 .}{phi}

Replacing $n$ with $k+\ell-1$, we have:

 \all{
 \phi^{k+\ell-1}_k(\theta)=\ST{k+\ell-1}{k}  \times A \Gamma(AG_{\theta}) \rho(\theta)
 .}{phi}

Comparing this with~\eqref{phi_def}, we find the solution for $\psi$ to be as follows: 
\all{
 \psi(k,\ell,\theta)=\ST{k+\ell}{k}   A \Gamma(AG_{\theta}) \rho(\theta)
 .}{psi_sol}

This readily yields $P(k,\ell,\theta)$,  using~\eqref{psi_def}. We get

\all{
P(k,\ell,\theta)=
\ST{k+\ell}{k}   A \Gamma(AG_{\theta}) \rho(\theta)
\fracc{\left(\frac{A\theta \beta_1}{\mu}\right)^k}{\Gamma (AG_{\theta}  +k+\ell +1)}
.}{Pfin1}

Plugging in the explicit expressions for $A$ and $G_{\theta}$, we arrive at the final solution:
\all{
P(k,\ell,\theta)=&
\ST{k+\ell}{k}  \left(\frac{\beta_1+\beta_2}{\beta_2}\right) \left(\frac{(\beta_1+\beta_2)  \beta_1}{  \beta_2} \fracc{\theta}{\mu}\right)^k  \nonumber \\ & \times
\fracc{\Gamma\left[\frac{\beta_1+\beta_2}{\beta_2}\left(1+\frac{\beta_1\theta}{\mu}\right)\right] }{\Gamma \left[\frac{\beta_1+\beta_2}{\beta_2}\left(1+\frac{\beta_1\theta}{\mu}\right) +k+\ell +1\right]}\rho(\theta)
.}{P_fin}

\section{Collapsed Joint Distribution of Degree and Fitness}
In real settings, the total number of links received by nodes are observed. For example, in the network of citations between scientific papers, what is observed and documented is $k+\ell$, that is, the total number of links (citations) received by papers. One cannot observe the number of citations that a paper receives purely based on its merit (meritocratic attachment),  or  the number of citations it receives due to its popularity (fame-driven attachment). This motivates us to derive the distribution of the total number of links, that is, $k+\ell$. Let us denote it by $q$. The joint distribution of $q,k$ is simply ${P(k,q-k,\theta)}$. If we sum over all possible values of $k$, we get:

 \all{
P(q,\theta)=&
\fracc{\left(\frac{\beta_1+\beta_2}{\beta_2}\right)\Gamma\left[\frac{\beta_1+\beta_2}{\beta_2}\left(1+\frac{\beta_1\theta}{\mu}\right)\right] }{\Gamma \left[\frac{\beta_1+\beta_2}{\beta_2}\left(1+\frac{\beta_1\theta}{\mu}\right) +q +1\right] }\rho(\theta)
 \nonumber \\ & \times
\sum_{k=0}^{q}
\ST{q}{k}  \left(\frac{(\beta_1+\beta_2)  \beta_1}{  \beta_2} \fracc{\theta}{\mu}\right)^{k} 
.}{Pq1}

From the properties of the unsigned Stirling numbers of the first kind, the sum on the right hand side of~\eqref{Pq1} can be readily evaluated. We have:

 \all{
P(q,\theta)=&
\fracc{\left(\frac{\beta_1+\beta_2}{\beta_2}\right)\Gamma\left[\frac{\beta_1+\beta_2}{\beta_2}\left(1+\frac{\beta_1\theta}{\mu}\right)\right] }{\Gamma\left( \frac{(\beta_1+\beta_2)  \beta_1}{  \beta_2} \fracc{\theta}{\mu}\right) }\rho(\theta)
 \nonumber \\ & \times
\fracc{\Gamma\left(q+ \frac{(\beta_1+\beta_2)  \beta_1}{  \beta_2} \fracc{\theta}{\mu}\right)}{\Gamma \left[q+1+\frac{\beta_1+\beta_2}{\beta_2}\left(1+\frac{\beta_1\theta}{\mu}\right)  \right]}
.}{Pq2}

\begin{figure}[h]
  \centering
  \includegraphics[width=.98 \columnwidth]{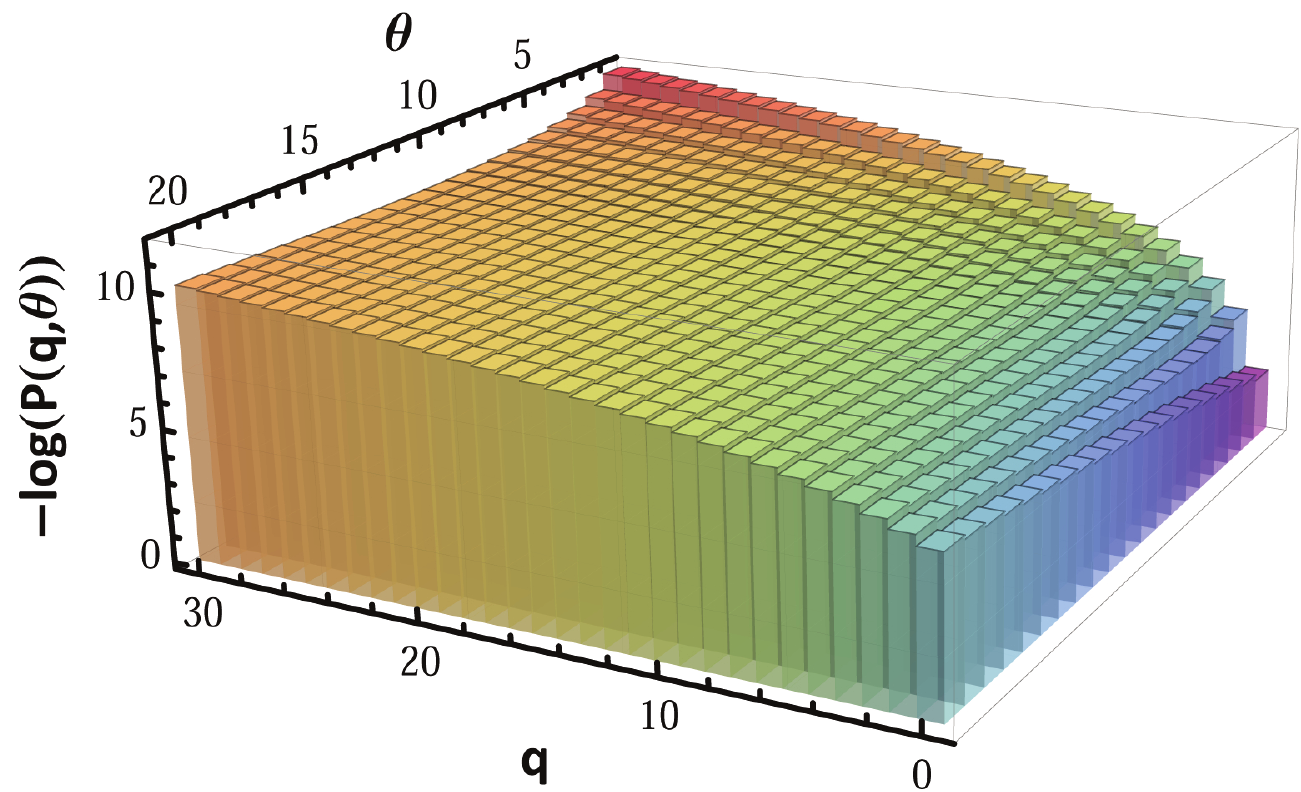}
  \caption[Figure ]%
  { Depiction of logarithm of inverse of $P(q,\theta)$. 
 We have considered an exponentially decaying fitness distribution for illustrative purposes:  $\rho(\theta)=0.9^{\theta}/10$. We have  also set ${\beta_1=2}$ and ${\beta_2=4}$. The logarithm is taken for smoothing purposes, due to the rapid plummet of the function $P(q,\theta)$ in the $q$ domain. 
} 
\label{pq3D}
\end{figure}

This is depicted in Figure~\ref{pq3D}.

If we use the generalization of binomial coefficients to non-integers, we can express~\eqref{Pq2} more concisely. Let us use the following notation for binomial coefficients: 
\all{
\BB{a}{b}=\fracc{\Gamma(a+1)}{\Gamma(b+1)\Gamma(a-b+1)},
}{binom}
where $a,b$ need not be integers. Using this notation, we can express~\eqref{Pq2} equivalently as follows: 
\all{
P(q,\theta)=\fracc{\beta_1+\beta_2}{\beta_1+2\beta_2}\fracc{\dbinom{\frac{\beta_1+\beta_2}{\beta_2}\left(1+\frac{\beta_1\theta}{\mu}\right)-1}{\frac{\beta_1+\beta_2}{\beta_2}}}{\dbinom{\frac{\beta_1+\beta_2}{\beta_2}\left(1+\frac{\beta_1\theta}{\mu}\right)+q}{\frac{\beta_1+\beta_2}{\beta_2}+1}} \rho(\theta)
.}{Pq_fin}

\begin{figure}[h]
  \centering
  \includegraphics[width=.9 \columnwidth]{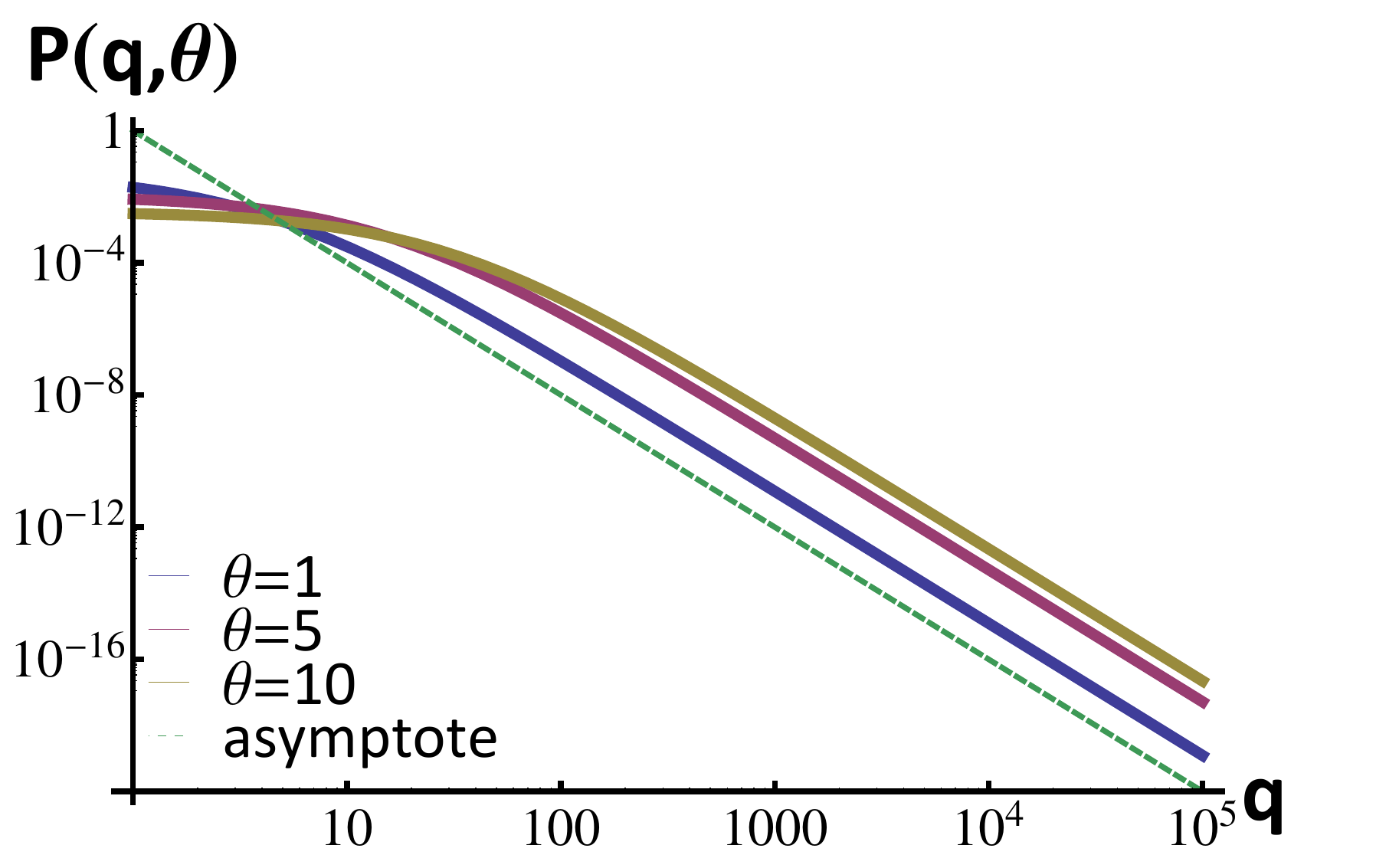}
  \caption[Figure ]%
  {The log-log plot of  $P(q,\theta)$ for three values of $\theta$ for exponential fitness distribution $\rho(\theta)=0.9^{\theta}/10$, with $\beta_1=8$ and $\beta_2=4$. As we expect from~\eqref{asym}, the curves are parallel for large values of $q$. It is visibly clear that the curves are also parallel to the asymptotic curve $q^{-2-\beta_1/\beta_2}$, as indicated by~\eqref{asym}. 
} 
\label{pqth}
\end{figure}

Now let us investigate the asymptotic behavior of $P(q,\theta)$ for large values of $q$. 
From~\eqref{Pq2}, we observe that only two terms have $q$. Using the Stirling approximation, we have
\all{
P(q,\theta) \sim \fracc{ (q+A G_{\theta})^{q+AG_{\theta}-\frac{1}{2}}e^{-(q+AG_{\theta})}}{(q+A G_{\theta}+A+1)^{q+AG_{\theta}+A+\frac{1}{2}}e^{-(q+AG_{\theta}+A+1)}}  
.}{asym1}

So we arrive at the following asymptotic relation

\all{
P(q,\theta) \sim q^{-2-\frac{\beta_1}{\beta_2}}~~~  \forall \theta
.}{asym}

Note that the exponent does not depend on $\theta$. This means that the degree distribution of the sub-populations with any fitness value follows the same exponent. In other words, the rate at which the degree distribution vanishes is the same for all fitness values. The  relative chances of different nodes attaining  extremely large degrees  depend  only on their fitness values, and not the degree itself, because if we divide the respective probabilities, only the fitness-dependent multiplicative factors would determine the ratio, as the $q$-dependent parts cancel out. This is illustrated in Figure~\ref{pqth}. Another implication of~\eqref{asym} is that the total degree distribution of the network, i.e. $P(q)$, has a power-law tail with exponent ${2+\frac{\beta_1}{\beta_2}}$.

\section{Expected Degree Distribution as a Function of Fitness}  

It is straightforward to compute the expected value of the degree distribution~\eqref{Pq2}. We have

 \all{
\langle q  \rangle_{\theta}=&\sum_{q=0}^{\infty}q   P(q|\theta )=\sum_{q=0}^{\infty}q \frac{ P(q,\theta)}{\rho(\theta)}
\nonumber \\ & =
\fracc{\left(\frac{\beta_1+\beta_2}{\beta_2}\right)\Gamma\left[\frac{\beta_1+\beta_2}{\beta_2}\left(1+\frac{\beta_1\theta}{\mu}\right)\right] }{\Gamma\left( \frac{(\beta_1+\beta_2)  \beta_1}{  \beta_2} \fracc{\theta}{\mu}\right) } 
 \nonumber \\ & \times
 \sum_{q=0}^{\infty} q
\fracc{\Gamma\left(q+ \frac{(\beta_1+\beta_2)  \beta_1}{  \beta_2} \fracc{\theta}{\mu}\right)}{\Gamma \left[q+1+\frac{\beta_1+\beta_2}{\beta_2}\left(1+\frac{\beta_1\theta}{\mu}\right)  \right]}
.}{qbar0}

We now perform the following summation: 
\all{
\mathcal{S} \stackrel{\text{def}}{=}
  \sum_{q=0}^{\infty} q
\fracc{\Gamma\left(q+ \frac{(\beta_1+\beta_2)  \beta_1}{  \beta_2} \fracc{\theta}{\mu}\right)}{\Gamma \left[q+1+\frac{\beta_1+\beta_2}{\beta_2}\left(1+\frac{\beta_1\theta}{\mu}\right)  \right]}
}{Sdef}

In Appendix~\ref{lem1},  we   prove the following identity for general real  positive  numbers  $y,x $:

\all{
\DD \sum_{q=0}^{\infty} \fracc{\Gamma(q+y)}{\Gamma(q+x+y)}=
\fracc{ \Gamma(y)} {(x-1)  \Gamma(y-1+x)}
.}{aux_iden}

%Now let us use this  identity~\eqref{aux} with ${y= \frac{(\beta_1+\beta_2)  \beta_1}{  \beta_2} \frac{\theta}{\mu}}$ and ${x=  \frac{\beta_1+\beta_2}{\beta_2}}$.   

If we use $x+1$ instead of $x$ in~\eqref{aux_iden}, we get
\all{
\DD \sum_{q=0}^{\infty} \fracc{\Gamma(q+y)}{\Gamma(q+x+y+1)}=
\fracc{ \Gamma(y)} {x  \Gamma(y +x)}
.}{yek}

Now note that, using the basic properties of the Gamma function, we can rewrite~\eqref{aux_iden} equivalently as follows
\all{
\DD \sum_{q=0}^{\infty} \fracc{ (q+x+y )\Gamma(q+y)}{\Gamma(q+x+y+1)}=
\fracc{ \Gamma(y)} {(x-1)  \Gamma(y-1+x)}
.}{do}

Expanding the left hand side, we have
 \all{
& \resizebox{.97\linewidth}{!}{$
 \DD \sum_{q=0}^{\infty} \fracc{ q\Gamma(q+y)}{\Gamma(q+x+y+1)}
+(x+y )
\DD \sum_{q=0}^{\infty} \fracc{ \Gamma(q+y)}{\Gamma(q+x+y+1)}
 $}
 \nonumber \\ & =
 \fracc{ \Gamma(y)} {(x-1)  \Gamma(y-1+x)}
 }{se}

Combining this with~\eqref{yek}, we arrive at
\all{
 \DD \sum_{q=0}^{\infty} \fracc{ q\Gamma(q+y)}{\Gamma(q+x+y+1)}
 = &
  \fracc{ \Gamma(y)} {(x-1)  \Gamma(y-1+x)}
  \nonumber \\ &
  -
 (x+y ) \fracc{ \Gamma(y)} {x  \Gamma(y +x)}
.}{chanar}

Using the basic properties of the Gamma function and algebraic simplifications, we can express this in the following form:
\all{
\resizebox{\linewidth}{!}{$
 \DD \sum_{q=0}^{\infty} \fracc{ q\Gamma(q+y)}{\Gamma(q+x+y+1)}
 =  
\fracc{\Gamma\left(y\right)}{\Gamma\left(x+y-1\right)}\fracc{y}{x \left(x-1\right)\left(x+y-1\right)}
.
$}
}{done}

This has the same form as~\eqref{Sdef}. We can use      identity~\eqref{done}  with ${y= \frac{(\beta_1+\beta_2)  \beta_1}{  \beta_2} \frac{\theta}{\mu}}$ and ${x=  \frac{\beta_1+\beta_2}{\beta_2}}$ to calculate $\mathcal{S}$ as follows:
\all{
\mathcal{S}=    &
 \fracc{\Gamma\left( \frac{(\beta_1+\beta_2)  \beta_1}{  \beta_2} \frac{\theta}{\mu}\right)} 
 {\Gamma\left(\frac{\beta_1 }{\beta_2}+ \frac{(\beta_1+\beta_2)  \beta_1}{  \beta_2} \frac{\theta}{\mu}\right)}
\nonumber \\ & \times 
\fracc{\left( \frac{(\beta_1+\beta_2)  \beta_1}{  \beta_2} \frac{\theta}{\mu} \right)}{\left(\frac{\beta_1+\beta_2}{\beta_2}\right)\left(\frac{\beta_1 }{\beta_2} \right)\left(\frac{\beta_1 }{\beta_2}+ \frac{(\beta_1+\beta_2)  \beta_1}{  \beta_2} \frac{\theta}{\mu} \right)}
 .}{S}
Plugging this into~\eqref{qbar0}, we get

\all{
\langle q  \rangle_{\theta}=(\beta_1+\beta_2)
 \frac{\theta  }{\mu} 
.}{qbar}

\begin{figure}[h]
  \centering
  \includegraphics[width=.9 \columnwidth]{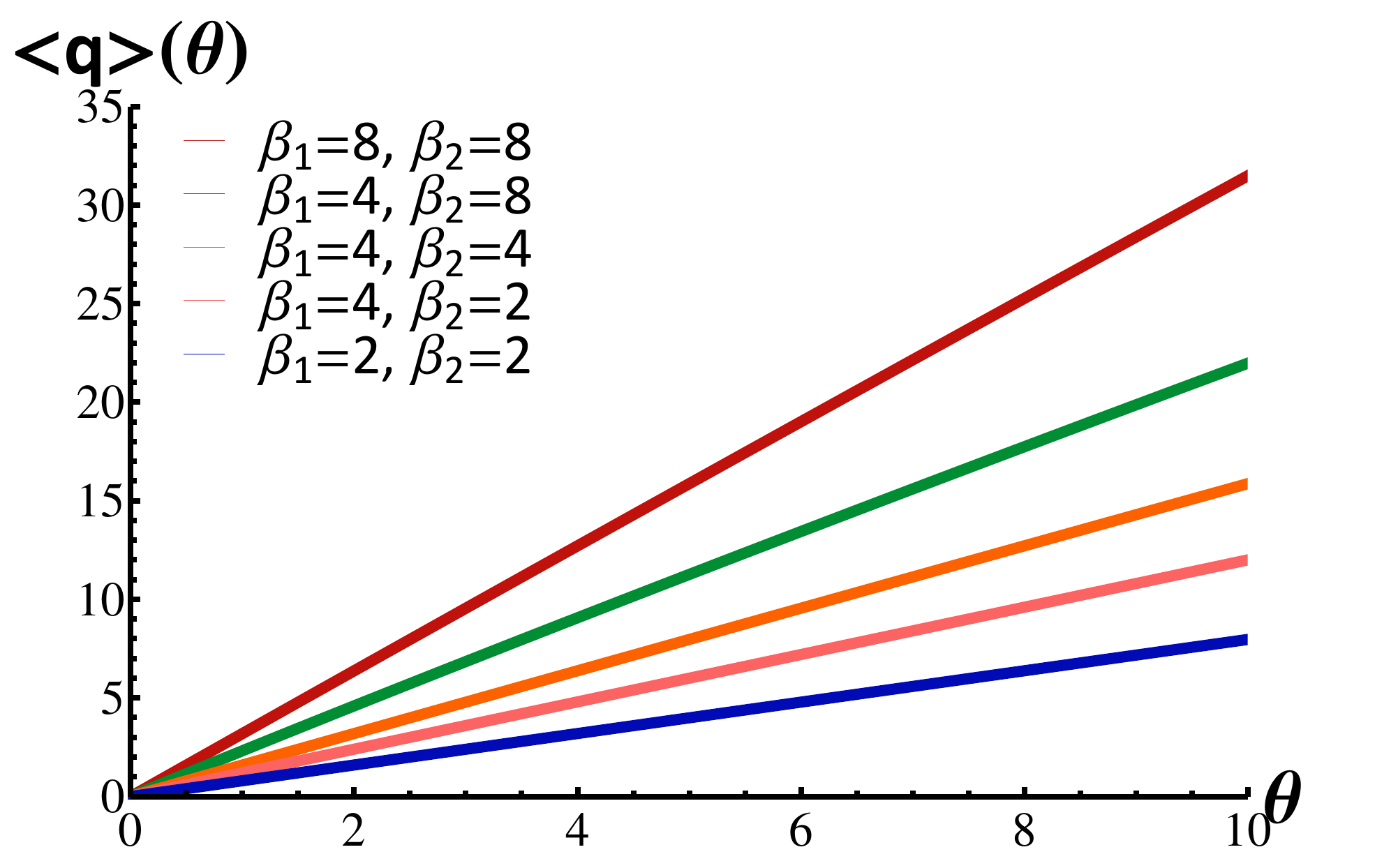}
  \caption[Figure ]%
  { As~\eqref{qbar} predicts, the expected values of $q$ is a linear function of $\theta$, with   slope   $\frac{\beta_1+\beta_2}{\mu}$ and    zero intercept. 
} 
\label{qbars}
\end{figure}

This is a linear relationship (see Figure~\ref{qbars}). 
If we take the average degree over all nodes, we need to sum up~\eqref{qbar} over all possible values of $\theta$. In the numerator, $\mu$ is created, which cancels out the $\mu$ in the denominator and we  get
\all{
\langle q  \rangle
=(\beta_1+\beta_2)
.}{qbar_trivial}
We know this result is true, because by construction, the total  number of links created in the system (which is always equal to the sum of in-degrees of all nodes) is $(\beta_1+\beta_2)t$ at  large times (when the effects of the initial conditions vanish) , and the total number of nodes is $t$, which means that their ratio (which yields the average degree) is equal to ${\beta_1+\beta_2}$.

%
%\begin{figure}[h]
%  \centering
%  \includegraphics[width=.9 \columnwidth]{lbars}
%  \caption[Figure ]%
%  { $\langle \ell  \rangle_{\theta}$ is an increasing function of $\beta_1$ and $\theta$, and is independent of $\beta_2$. This manifests the coupling between layers:  $\theta$ resides solely in layer 1, yet it affects $\ell$ in the other layer. 
%%This independence is expected because the growth mechanism in layer 1 is not coupled to layer 2, and the dynamical dependence is unilateral. 
%} 
%\label{kbars}
%\end{figure}

%
%
%\begin{figure}[h]
%  \centering
%  \includegraphics[width=.9 \columnwidth]{lbars}
%  \caption[Figure ]%
%  { $\langle \ell \rangle_{\theta}$ is an increasing function of $\beta_2$ and $\theta$, as expected. 
%} 
%\label{lbar}
%\end{figure}

\section{Summary and Open Problems }

This paper extends the literature of multiplex networks by introducing a simple model which incorporates intrinsic fitness \emph{and} preferential attachment. The merit layer is latent, yet drives the growth mechanism and the degree dynamics. We obtained closed-form expressions for the joint  interlayer distribution of degrees and fitness, as well as that of the total degrees. We observed that the expected value of the total degree linearly increases with fitness. 

An immediate generalization of the present problem would be its extension to an  arbitrary number of layers. Also, we have disregarded the temporal dynamics of the system and its transients in favor of the steady state. This loses valuable information about the transient state and the effects of initial conditions on the evolution of the network. In other words, for a given initial network (not  necessarily small), one can study the evolution of the system in arbitrary time regimes, and investigate how the properties of the initial network affect the equilibration of the system, and how they affect the asymptotic properties of the network. 

Another immediate step to augment the present model is to devise statistical recipes for inference. Since fitness is a latent variable and only $q$ can be observed, one can use~\eqref{Pq2} (or its time-dependent version)  to devise maximum likelihood techniques to infer the distribution of fitness (merit) of scientific publications, blog posts, etc., by observing the distribution (or evolution)  of   degrees.

\appendix 
\section{Proof of Identity~\eqref{aux_iden}}\label{lem1}

We need to prove the following identity 

\all{
\DD \sum_{q=0}^{\infty} \fracc{\Gamma(q+y)}{\Gamma(q+x+y)}=
\fracc{ \Gamma(y)} {(x-1)  \Gamma(y-1+x)}
.}{a1}

The definition of the Beta function for positive real values $x,y$ is
\all{
B(a,b)=\DD \int_0^1 t^{a-1} (1-t)^{b-1} dt= \fracc{\Gamma(a) \Gamma(b)}{\Gamma(a+b)}
.}{beta}

We can rewrite the summand of~\eqref{a1} as follows:
\all{
\fracc{\Gamma(q+y)}{\Gamma(q+x+y)}=
\fracc{\DD \int_0^1 t^{q+y-1} (1-t)^{x-1} dt}{\Gamma(x)}
.}{a2}

We now have
\all{
&
\DD \sum_{q=0}^{\infty} \fracc{\Gamma(q+y)}{\Gamma(q+x+y)}
=\fracc{1}{\Gamma(x)}\DD \sum_{q=0}^{\infty}  \DD \int_0^1 
dt  (1-t)^{x-1}
 t^{q+y-1}
 \nonumber \\ &
=
 \fracc{1}{\Gamma(x)}\DD  \DD \int_0^1 
dt  (1-t)^{x-1} t^{y-1} \sum_{q=0}^{\infty}
t^q
 \nonumber \\ &
 =
 \fracc{1}{\Gamma(x)}\DD  \DD \int_0^1 
dt  (1-t)^{x-1} t^{y-1} \frac{1}{1-t}
 \nonumber \\ &
 =
 \fracc{1}{\Gamma(x)}\DD  \DD \int_0^1 
dt  (1-t)^{x-2} t^{y-1}  
 \nonumber \\ &
 =
 \fracc{1}{\Gamma(x)}B(  x-1 , y  ) =  \fracc{1}{\Gamma(x)} \fracc{\Gamma(x-1) \Gamma(y)}{\Gamma(x+y-1)}
  \nonumber \\ &
 =
  \fracc{1}{(x-1)\Gamma(x-1)} \fracc{\Gamma(x-1) \Gamma(y)}{\Gamma(x+y-1)}
    \nonumber \\ &
 =
\fracc{ \Gamma(y)} {(x-1)  \Gamma(y-1+x)}
,}{a3}
which concludes the proof.

\bibliographystyle{apsrev4-1}
\bibliography{myref}

\end{document}